\documentclass[preprint,12pt]{article}
\usepackage{amssymb}
\usepackage{graphicx}
\usepackage{multirow}
\usepackage{caption}
\usepackage{subfig}
\begin{document}
\title
{Imprint of modified Einstein's gravity on white dwarfs: Unifying type~Ia supernovae}
\author{Upasana Das and Banibrata Mukhopadhyay\\ \\
Department of Physics, Indian Institute of
  Science, Bangalore 560012, India\\
upasana@physics.iisc.ernet.in , bm@physics.iisc.ernet.in}
\maketitle
\begin{center}
Essay received Honorable Mention in the Gravity  Research Foundation \\ 2015 Awards 
for Essays on Gravitation
\end{center}

\vskip1.0cm
\begin{abstract}
We establish the importance of modified Einstein's gravity (MG) in white dwarfs (WDs) for the first time 
in the literature. 
We show that MG leads to significantly sub- and super-Chandrasekhar limiting mass WDs, 
depending on a single model parameter. However, conventional WDs on approaching Chandrasekhar's 
limit are expected to trigger type Ia supernovae (SNeIa), a key to unravel the evolutionary 
history of the universe. Nevertheless, observations of several peculiar, under- and over-luminous 
SNeIa argue for the limiting mass widely different from Chandrasekhar's limit. 
Explosions of MG induced sub- and super-Chandrasekhar limiting mass WDs  
explain under- and over-luminous SNeIa respectively, thus unifying these
two apparently disjoint sub-classes. Our discovery questions both the global validity of Einstein's gravity 
and the uniqueness of Chandrasekhar's limit.
\end{abstract} 


\newpage

\noindent \textbf{Introduction}
\vskip0.5cm

\noindent 
The validity of Einstein's theory of general relativity (GR) has been tested extensively in the weak field regime,
e.g., through laboratory experiments and solar-system tests. The question is, whether GR is the ultimate theory of
gravitation, or it requires modification in the strong gravity regime. 
Indeed, it was shown that modified gravity (MG) theories reveal significant deviations to
the GR solutions for neutron stars (NSs) \cite{dam}.
As NSs are much more compact than white dwarfs (WDs), so far, MG theories
have been applied only to them in order to test the validity of such theories in the strong
field regime. The current venture is to show that a MG theory is indispensable in the context of WDs, which is a first in 
the literature to the best of our knowledge. The motivation is the following.

Type~Ia supernovae (SNeIa) are believed to result from the
violent thermonuclear explosion of a carbon-oxygen WD,
when its mass approaches the famous Chandrasekhar limit of
$1.44M_\odot$, when $M_\odot$ is the solar mass. SNIa is used as a standard candle in understanding
the expansion history of the universe \cite{perl99}. 

However, some of these SNeIa are highly over-luminous, e.g.  SN 2003fg, SN 2006gz, SN 2007if, SN 2009dc \cite{howel,scalzo},
and some others are highly under-luminous, e.g. SN 1991bg, SN 1997cn, SN 1998de, SN 1999by
\cite{1991bg,taub2008}. 
The luminosity of the former group (super-SNeIa) implies highly super-Chandrasekhar 
WDs, having mass $2.1-2.8M_\odot$, as their most plausible progenitors \cite{howel,scalzo}. 
While, the latter group (sub-SNeIa) predicts the progenitor mass could be as low
as $ \sim M_\odot$ \cite{1991bg}.  
Moreover, the characteristic lightcurves of these SNeIa are quite 
peculiar compared to their conventional counterparts.
The models attempted to explain them so far entail caveats.

A major concern, however, is that a large array
of models is required to explain apparently the same phenomena, i.e., triggering of thermonuclear explosions
in WDs. It is unlikely that nature would seek mutually antagonistic scenarios to exhibit sub- and super-SNeIa.
This is where the current work steps in, which unifies
the phenomenologically disjoint sub-classes of SNeIa by a single underlying theory. 
This is achieved by invoking 
a MG theory in WDs.

\vskip1.0cm

\noindent \textbf{Basic equations and modified gravity model}
\vskip0.5cm

\noindent 
Let us start with the 4-dimensional action as \cite{livrel}
\begin{equation}
S = \int \left[\frac{1}{16\pi} f(R) + {\cal L}_M \right] \sqrt{-g}~ d^4 x , 
\end{equation}
where $g$ is the determinant of the metric tensor $g_{\mu\nu}$, 
${\cal L}_M$ the Lagrangian density of the matter field, 
$R$ the scalar curvature defined as $R=g^{\mu\nu}R_{\mu\nu}$, 
when $R_{\mu\nu}$ is the Ricci tensor and $f$ is an 
arbitrary function of $R$ (in GR $f(R)=R$). For the present purpose, we choose the Starobinsky model 
\cite{starobi} defined as $f(R)=R+\alpha R^2$, when $\alpha$ is a constant. 
However, similar effects could also be obtained in other MG theories, e.g. 
Born-Infeld gravity (e.g. \cite{bana}). Now, on extremizing the above action 
one obtains the modified field equation as
\begin{equation}
G_{\mu\nu}+\alpha X_{\mu\nu}= 8\pi T_{\mu\nu},
\label{modfld}
\end{equation}
where $G_{\mu\nu}$ is Einstein's field tensor, $T_{\mu\nu}$ is the energy-momentum tensor of the matter field 
and $X_{\mu\nu}$ is a function of $g_{\mu\nu}$, $R_{\mu\nu}$ and $R$.

\vskip1.0cm
\noindent \textbf{Solution procedure}
\vskip0.5cm

\noindent 
In order to solve equation (\ref{modfld}), we adopt the 
perturbative method (e.g. \cite{capo1}), 
such that $\alpha R<<1$. Further, we consider the hydrostatic equilibrium condition: $g_{\nu r}\nabla_\mu T^{\mu\nu}=0$, 
with zero velocity and $\nabla_\mu$ the covariant derivative. Hence, we 
obtain the differential equations for mass $M_\alpha (r)$, pressure $P_\alpha (r)$
(or density $\rho_\alpha (r)$)
and gravitational potential $\phi_\alpha (r)$, of spherically symmetric WDs, where $r$ is the 
radial coordinate: the {\it modified} Tolman-Oppenheimer-Volkoff (TOV) equations. 
For $\alpha=0$, these  equations reduce to TOV equations in GR.

In order to solve the modified TOV equations, we must supplement them with 
appropriate boundary conditions and an equation of state (EoS) obtained 
by Chandrasekhar \cite{chandra35}, given by $P_\alpha=K\rho_\alpha^{1+(1/n)}$, for 
extremely low and high densities, where $n$ is the polytropic index and $K$ a dimensional constant. 
The boundary conditions are $M_\alpha (0)=0$ and $\rho_\alpha (0) =\rho_c$, where $\rho_c$ is the central 
density of the WD. Note that a particular $\rho_c$ corresponds to a particular mass $M_*$ and radius $R_*$ 
of WDs. Hence, by varying $\rho_c$ from $2\times 10^5$ gm/cc to $10^{11}$ gm/cc, we construct the mass-radius 
relation.

\vskip1.0cm
\noindent \textbf{Results}
\vskip0.5cm

\noindent 
Figures 1(a) and (b) show that, for $\alpha=0$ (GR case), with the increase of $\rho_c$, 
$M_*$ increases and $R_*$ decreases, until it reaches a maximum mass $M_{\rm max}=1.405M_\odot$ (smaller 
than the Newtonian Chandrasekhar limit of $1.44M_\odot$)
at $\rho_c=3.5\times 10^{10}$ gm/cc. 
A further increase in $\rho_c$ results in a slight decrease in $M_*$, indicating the onset of an unstable branch.

\begin{figure}[h]
\begin{center}
\includegraphics[angle=0,width=18cm]{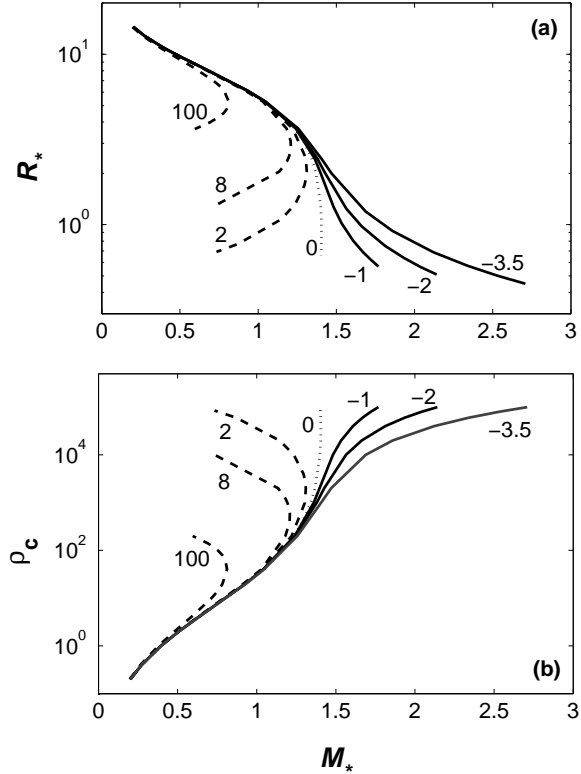}
\caption{Unification diagram for SNeIa:
(a) Mass-radius relations. (b) Variation of $\rho_c$ with $M_*$.
The numbers adjacent to the various lines denote 
$\alpha/(10^{13}~{\rm cm^2})$. 
$\rho_c$, $M_*$ and $R_*$ are in units of $10^6$ gm/cc, $M_\odot$ and 1000 km, respectively.
  }
\label{fig}
\end{center}
\end{figure}

Coming to the $\alpha<0$ cases, Figure 1(b) shows that for $\rho_c>10^8$ gm/cc, the 
$M_*-\rho_c$ curves deviate from the GR curve due to MG effects. 
This reveals that MG has a tremendous impact on WDs which so far was completely overlooked.
Note that $M_{\rm max}$ for all the three cases corresponds to $\rho_c=10^{11}$ gm/cc, 
an upper-limit chosen to avoid possible neutronization. 
Interestingly, all values of $M_{\rm max}$ are highly super-Chandrasekhar, ranging from 
$1.8-2.7M_\odot$. 
Thus while the GR effect is very small, MG effect could
lead to $\sim 100\%$ increase in the limiting mass of WDs. 
The corresponding values of $\rho_c$ are large enough to initiate thermonuclear reactions, e.g.
they are larger than $\rho_c$ corresponding to $M_{\rm max}$ of $\alpha=0$ case, whereas the respective 
core temperatures are expected to be similar. This 
explains the entire range of the observed super-SNeIa mentioned above \cite{howel,scalzo}.


Table 1 ensures the perturbative validity of the solutions. Recall that 
we solve the modified TOV equations only up to ${\cal O} (\alpha)$. Since the product 
$\alpha R$ is first order in $\alpha$, we replace $R$ in it by the zeroth order Ricci scalar 
$R^{(0)} = 8\pi(\rho^{(0)} - 3P^{(0)})$, where $\rho^{(0)}$ and $P^{(0)}$ are the zeroth order solutions of density 
and pressure respectively, obtained in GR (when $\alpha=0$). For the perturbative validity of the entire solution, 
$|\alpha R^{(0)}|_{\rm max} \ll 1$ should hold 
true. Next we consider $g_{tt}^{(0)}/g_{tt}$ and $g_{rr}^{(0)}/g_{rr}$ (ratios of $g_{\mu\nu}$-s 
in GR and those in MG up to ${\cal O} (\alpha)$),
which should be close to 1 for the validity of perturbative method \cite{orelana}. Hence,
$|1-g_{tt}^{(0)}/g_{tt}|_{\rm max} \ll 1$ and 
$|1-g_{rr}^{(0)}/g_{rr}|_{\rm max} \ll 1$ should both hold true.
Table 1 shows that all the three measures quantifying perturbative validity are at least $2-3$ 
orders of magnitude smaller than 1.

Coming to the $\alpha>0$ cases, Figure 1(b) shows 
that all the three $M_*-\rho_c$ curves overlap with the $\alpha=0$ curve in the low density region. 
However, with the increase 
of $\alpha$, the region of overlap recedes to a lower $\rho_c$. MG 
effects set in at $\rho_c \gtrsim 10^8,~4\times 10^7$ and $2\times 10^6$ gm/cc, 
for $\alpha = 2\times 10^{13}~ {\rm cm^2}$, $8\times 10^{13}~ {\rm cm^2}$ and 
$10^{15}~{\rm cm^2}$ respectively. For a given $\alpha$, with the increase of $\rho_c$, $M_*$ 
first increases, reaches a maximum and then decreases, like the $\alpha=0$ case. With the 
increase of $\alpha$, $M_{\rm max}$ decreases and, interestingly,
for $\alpha=10^{15}~{\rm cm}^2$, it is highly sub-Chandrasekhar ($0.81M_\odot$).
In fact, $M_{\rm max}$ for all the chosen $\alpha>0$ is sub-Chandrasekhar, ranging $1.31-0.81M_\odot$. 
This is a remarkable finding since it establishes that even if $\rho_c$-s for 
these sub-Chandrasekhar WDs are lower than the conventional value at which SNeIa are usually triggered, an attempt 
to increase the mass beyond $M_{\rm max}$ with increasing $\rho_c$, for a given $\alpha$, will lead to a gravitational instability. 
This presumably will be followed by a 
runaway thermonuclear reaction, provided the core temperature increases sufficiently due to collapse. 
Occurrence of such thermonuclear runway reactions, triggered at densities as low as $10^6$ gm/cc,
has already been demonstrated \cite{runaway}. Thus, once $M_{\rm max}$ is approached, a SNIa is expected to trigger just 
like in the $\alpha=0$ case, explaining the sub-SNeIa \cite{1991bg,taub2008}, 
like SN 1991bg mentioned above. 
Table 2 confirms that the solutions for the $\alpha>0$ cases are within the perturbative regime.

\begin{table}[]
\caption{Measure of validity of perturbative solutions for $\alpha<0$ corresponding to $M_{\rm max}$ in Fig. \ref{fig}.}
\begin{center}
\small
\begin{tabular}{|c|c|c|c|c|c|c|c|}

\hline 
$\alpha/(10^{13}~{\rm cm}^2)$  & 
$|\alpha R^{(0)}|_{\rm max}$ & $|1-g^{(0)}_{tt}/g_{tt}|_{\rm max}$ & $|1-g^{(0)}_{rr}/g_{rr}|_{\rm max}$ \\ 
\hline

-1 & 0.00184 & 0.0016 & 0.0052 \\ 

-2 &  0.00369 & 0.0031 & 0.0108 \\



-3.5 & 0.00646 & 0.0052 & 0.0199 \\ \hline

\end{tabular}

\end{center}

\end{table}

\begin{table}[]
\caption{Measure of validity of perturbative solutions for $\alpha>0$ corresponding to $M_{\rm max}$ in Fig. \ref{fig}.}
\begin{center}
\small
\begin{tabular}{|c|c|c|c|}

\hline 
$\alpha/(10^{13}~{\rm cm}^2)$  & $|\alpha R^{(0)}|_{\rm max}$ & $|1-g^{(0)}_{tt}/g_{tt}|_{\rm max}$ & $|1-g^{(0)}_{rr}/g_{rr}|_{\rm max}$ \\ 
\hline

2 & $7.4\times 10^{-5}$ &  $6.8\times 10^{-5}$ & $2.0\times 10^{-4}$ \\ 

8 & $7.4\times 10^{-5}$ & $6.8\times 10^{-5}$ & $2.0\times 10^{-4}$ \\

100 & $7.4\times 10^{-5}$ & $6.9\times 10^{-5}$ & $2.0\times 10^{-4}$  \\ \hline

\end{tabular}

\end{center}

\end{table}

\newpage
\vskip1.0cm
\noindent \textbf{Conclusions}
\vskip0.5cm

\noindent 
Based on a simple $f(R)$-model, we show that modifications to GR
are indispensable in WDs, especially for determining their limiting mass. It remarkably explains and 
unifies a wide range of observations for which GR is insufficient. 
We note here that the perturbative method is adequate for 
the present study, as then we have a handle on $\alpha$ characterizing our model which cannot be 
arbitrarily large, allowing it to be constrained directly by astrophysical observations \cite{naf}.
Hence, depending on the magnitude and sign of $\alpha$, we not only obtain both 
highly super-Chandrasekhar (for $\alpha<0$) and highly sub-Chandrasekhar (for $\alpha>0$) 
limiting mass WDs, but we also establish them as progenitors of the peculiar, over-luminous and 
under-luminous SNeIa, respectively. 
We further note that even though $\alpha$ is assumed to be constant within individual WDs, there is indeed 
an implicit dependence of $\alpha$ on $\rho_c$, as evident from Figure 1(b), indicating the existence of 
an underlying chameleon effect \cite{chameleon}. 
Thus, a single underlying theory, 
inspired by the need to modify Einstein's 
theory of GR, unifies the two apparently disjoint sub-classes of SNeIa, 
which have so far hugely puzzled astronomers. 
The significance of the current work lies in the 
fact that it not only questions the uniqueness of the Chandrasekhar mass-limit for WDs, 
but it also argues for the need of a {\it modified theory} of GR to explain the 
observable universe.

\end{document}